\RequirePackage{filecontents}

\documentclass[reprint,12pt]{elsarticle}

\usepackage{graphicx}               
\usepackage[caption=false]{subfig}  
\usepackage{amsmath, amssymb}       
\usepackage{wasysym}
\usepackage{lmodern}                
\usepackage[T1]{fontenc}            
\usepackage[utf8]{inputenc}  
\usepackage{indentfirst}            
\usepackage{physics}
\usepackage{subcaption}
\usepackage{natbib}

\begin{document}

\begin{frontmatter}

\title{Escape and transport in chaotic motion of charged particles in a magnetized plasma under the influence of two and three modes of drift waves}

\author[UFPR]{P. Haerter\corref{teste}}
\cortext[teste]{haerter@ufpr.br}
\author[UFPR,UFPR2]{R.L. Viana}
\author[UNESP]{E. D. Leonel}

\affiliation[UFPR]{organization={Departamento de Física, Universidade Federal do Paraná},
            city={Curitiba},
            postcode={81531-990}, 
            state={Paraná},
            country={Brazil}}
      
\affiliation[UFPR2]{organization={Centro Interdisciplinar de Ciência, Tecnologia e Inovação, Núcleo de Modelagem e Computação Científica, Universidade Federal do Paraná},
            city={Curitiba},
            postcode={81531-990}, 
            state={Paraná},
            country={Brazil}}

\affiliation[UNESP]{organization={Departamento de Física, Universidade Estadual Paulista},
            city={Rio Claro},
            postcode={13506-900}, 
            state={São Paulo},
            country={Brazil}}

\begin{abstract}
This study investigates how two- and three-wave configurations govern particle escape and transport in tokamak edge plasmas. Using a Hamiltonian model derived from drift-wave turbulence, we analyze test particle dynamics through Poincaré maps, fractal escape basins, and entropy metrics. Introducing a third wave increases basin entropy, enhancing particle escape rates while reducing basin boundary entropy, indicative of suppressed basin mixing. Escape time analyses reveal resonant scattering disrupts coherent transport pathways, linking fractal absorption patterns to heat load mitigation in divertors. Characteristic transport is also analyzed and regimes transition between anomalous $(\alpha > 1)$ and normal diffusion $(\alpha \approx 1)$, two-wave systems sustain anomalous transport, while the third wave homogenizes fluxes through stochastic scattering. Fractal structures in escape basins and entropy-driven uncertainty quantification suggest strategies to engineer transport properties, balancing chaos and order for optimized confinement.   
\end{abstract}


\end{frontmatter}

\section{Introduction}

The confinement of high-temperature plasmas in magnetic fusion devices remains a critical challenge due to the interplay of turbulent fluctuations and anomalous particle transport \cite{diamondZonalFlowsPlasma2005,chenIntroductionPlasmaPhysics2016}. Drift waves, driven by density and temperature gradients in magnetized plasmas, generate turbulent electric fields that induce chaotic particle motion. This motion results in particle escape rates that far exceed predictions from classical collisional theory. Such anomalous transport degrades plasma performance and risks damaging reactor walls through localized heat and particle fluxes, necessitating advanced mitigation strategies \cite{balescuAspectsAnomalousTransport2005,hortonDriftWavesTransport1999}.  

Particle transport in high-temperature plasmas is dominated by long-range electric fields arising from Coulomb interactions among charged particles. In magnetized plasmas, low-frequency ion acoustic oscillations emerge as critical collective modes, governing transport properties such as resistivity and thermal conductivity. Steep density and temperature gradients trigger instabilities in these modes, leading to drift-wave turbulence and consequent anomalous transport of particles and energy.  

Among the various instabilities in tokamaks, electrostatic drift-wave turbulence in the plasma edge plays a pivotal role in governing cross-field transport \cite{diamondZonalFlowsPlasma2005}. These low-frequency fluctuations, driven by steep density and temperature gradients, generate turbulent electric fields that drive $\mathbf{E} \times \mathbf{B}$ drifts—a key mechanism for chaotic particle motion \cite{hasegawaPseudothreedimensionalTurbulenceMagnetized1978}.  

A typical framework for analyzing these dynamics is the Horton model\cite{hortonOnsetStochasticityDiffusion1985}, which reduces the plasma edge to a slab geometry with a uniform magnetic field \(\mathbf{B} = B\hat{z}\). Particle motion is described through \(\mathbf{E}\times\mathbf{B}\) drifts governed by a Hamiltonian formulated from electrostatic potentials. By decomposing the potential into an equilibrium component (\(\Phi_0(x)\)) and perturbative contributions from \(N\) drift waves, this model predicts the emergence of fractal structures in phase space, where chaotic and quasi-regular trajectories intertwine.  

Recent studies \cite{mathiasFractalStructuresChaotic2017,vianaFractalBoundariesChaotic2017} demonstrate how drift-wave turbulence generates fractal escape basins—regions of initial conditions leading to particle loss—characterized by self-similar patterns and Wada boundaries. In these boundaries, three or more basins share a common fractal edge, a feature critical for understanding particle migration toward plasma-facing components such as divertors. Here, fractal absorption patterns directly influence heat and particle load distributions \cite{mathiasFractalStructuresMagnetic2022,dasilvaEscapePatternsMagnetic2002}, complicating efforts to optimize reactor durability. The transition from integrable to chaotic dynamics under varying wave configurations further challenges confinement, as resonant interactions between modes amplify stochasticity and disrupt coherent transport pathways.

Quantitative tools such as basin entropy $(S_b)$ and basin boundary entropy $(S_{bb})$\cite{dazaBasinEntropyNew2016} have emerged to characterize the uncertainty and fractality of escape processes. These metrics reveal how perturbations amplify unpredictability in particle trajectories, even in systems with seemingly uniform chaotic regions. For instance, in tokamaks with ergodic magnetic limiters, fractal basins persist despite attempts to homogenize heat loads, highlighting the intrinsic complexity of plasma-wall interactions \cite{mathiasFractalStructuresChaotic2017,mathiasFractalStructuresMagnetic2022,souzaBasinEntropyShearless2023,haerterBasinEntropyWada2023}.  Reduced fractal dimensions correlate with elevated anomalous diffusion exponents, suggesting plasma edge turbulence alters transport via analogous mechanisms \cite{balescuAspectsAnomalousTransport2005}.  

This paper extends prior work by analyzing how two- and three-wave configurations alter escape and transport regimes in magnetized plasmas. The manuscript is structured as follows: Section~\ref{sec:drift_waves_model} outlines the theoretical model and numerical framework; Section~\ref{sec:escape_basins} details particle escape dynamics and its impact on trajectory uncertainty across perturbation ranges; Section~\ref{sec:particle_transport} presents transport analyses across periodic and chaotic regimes, contrasting diffusion behaviors under varying wave amplitudes; and Section~\ref{sec:conclusions} summarizes key conclusions and implications for plasma confinement.  

\section{Drift-Waves model}
\label{sec:drift_waves_model}

\begin{figure*}[h!]
    \centering
    \includegraphics[scale=0.2]{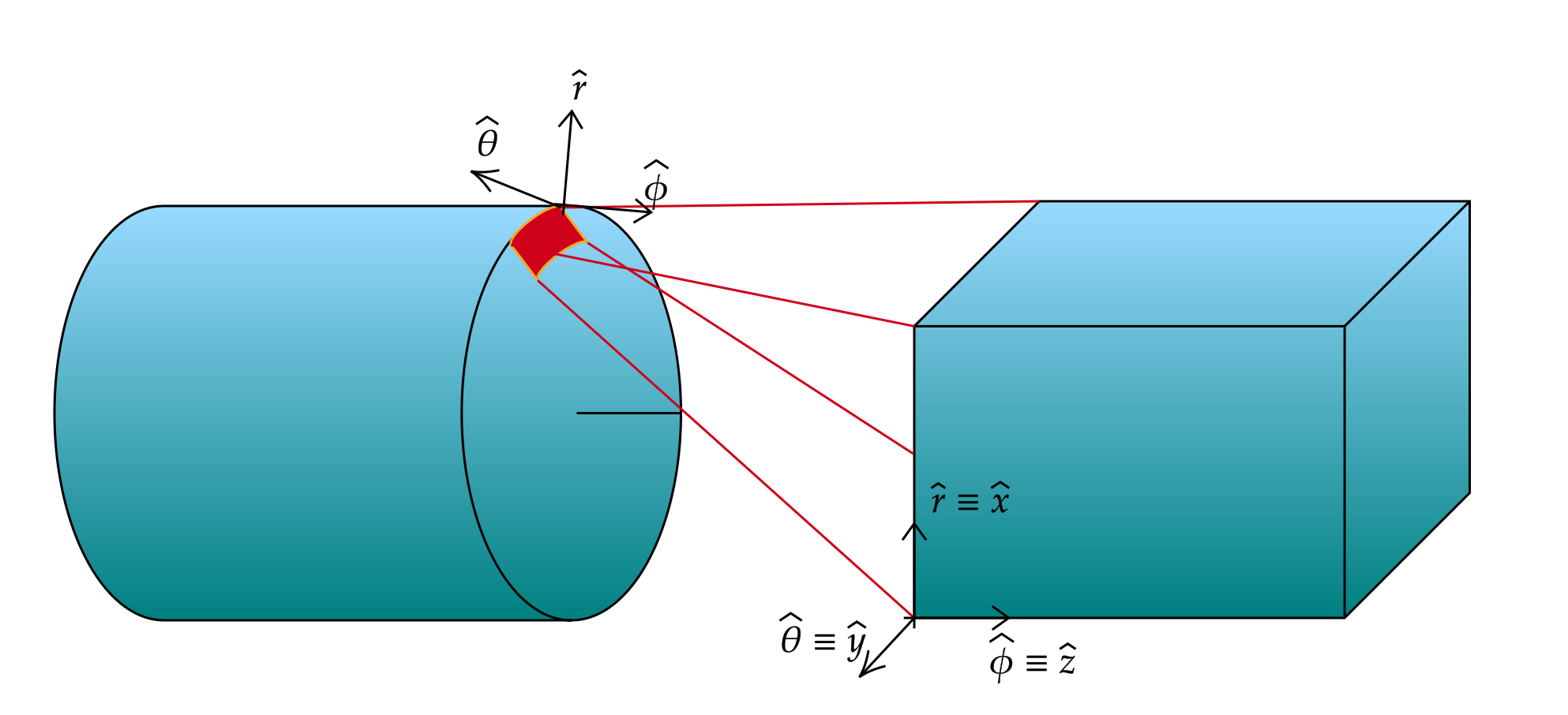}
    \caption{Graphical representation of the transformation of cylindrical coordinates to rectangular ones.}
    \label{fig:Diagrama}
\end{figure*}

A well-known model for describing the influence of electrostatic potential turbulence caused by drift waves in the plasma edge is the one proposed by Horton \cite{hortonOnsetStochasticityDiffusion1985}. Neglecting curvature effects, the plasma edge region can be approximated by a slab geometry. We therefore transform the toroidal coordinates $(r, \theta, \phi)$ into a Cartesian system $(x, y, z)$, as illustrated in Fig.~\ref{fig:Diagrama}. Here, $x$ corresponds to the radial direction, $y$ represents the poloidal direction (with periodicity $2\pi$), and $z$ denotes the toroidal direction. The uniform magnetic field is expressed as $\vb{B} = B\hat{z}$.

We assume that the ion and electron charges do not appreciably affect the electric and magnetic fields; hence, the corresponding guiding centers can be considered as passive tracers, or test particles. Such test particles move with the $\vb{E} \cross \vb{B}$ drift velocity.

\begin{equation}
    \vb{v} = - \frac{\nabla \Phi \cross \vb{B}}{B^2},
\end{equation}
where the electric field is written as $\vb{E} = -\nabla \Phi$, with $\Phi(x,y,t)$ being a scalar potential. This system is equivalent to the following set of canonical equations:
\begin{align}
    v_x &= \dv{x}{t} = -\frac{1}{B_0}\pdv{\Phi}{y}, \label{eq:vx} \\
    v_y &= \dv{y}{t} = \frac{1}{B_0}\pdv{\Phi}{x}, \label{eq:vy}
\end{align}
corresponding to the time-dependent Hamiltonian
\begin{equation}
    H(x,y,t) = \frac{1}{B_0}\Phi(x,y,t).
\end{equation}

The electric potential can be divided into two parts: an equilibrium part $\Phi_0(x)$ corresponding to a radial electric field, and a perturbation caused by $N$ drift waves with amplitudes $A_i$, frequencies $\omega_i$, and wave vectors $\vb{k}_i = (k_{x,i}, k_{y,i})$ for $i = 1, 2, \ldots, N$. This gives:
\begin{equation}
    \Phi(x, y, t) = \Phi_0(x) + \sum_{i=1}^{N} A_i \sin(k_{x,i} x) \cos(k_{y,i} y - \omega_i t),
    \label{eq:Horton-phi}
\end{equation}
where we assume a stationary wave pattern along the radial direction $x$ and a traveling wave along the poloidal direction $y$. Let $L_x$ and $L_y$ be the characteristic lengths along these directions, such that:
\begin{equation}
    k_x = \frac{n \pi}{L_x}, \qquad k_y = \frac{2 \pi m}{L_y},
\end{equation}
where $m$ and $n$ are suitably chosen positive integers.

In the case of $N$ drift waves, the Hamiltonian reads:
\begin{equation}
\begin{aligned}
    H(x,y,t) &= \frac{1}{B_0}[\Phi_0(x) + \\
    &\sum_{i=1}^{N} A_i \sin(k_{x,i} x) \cos(k_{y,i} y - \omega_i t)].
\end{aligned}    
\end{equation}

After transitioning to a reference frame moving with the phase velocity of the first wave, $u_1$, and considering the resonant case where the drift velocity satisfies 
$$
v_e = \frac{1}{B}\dv{\Phi_0}{x} = u_1 = \frac{\omega_1}{k_{y,1}},
$$
the Hamiltonian becomes:
\begin{equation}
\begin{aligned}
    H(x,y,t) = A_1 \sin(k_{x,1} x)\cos(k_{y,1} y) + \\\sum_{i=2}^{N} A_i \sin(k_{x,i} x + \beta_i) \cos\big(k_{y,i}(y - u_i t)\big),
    \label{eq:H_Final}
\end{aligned}    
\end{equation}

where $u_i = \frac{\omega_i}{k_{y,i}} - \frac{\omega_1}{k_{y,1}}$ and $\beta_i$ is a phase term associated with the breakdown of transport barriers in symmetry regions. This phase enables particle transport along the $x$-direction, which is the focus of this study\cite{klevaStochasticExBParticle1984,bosioIdentifyingBallisticModes2025}.

\begin{figure*}[h!]
    \centering
    \includegraphics[scale=0.5]{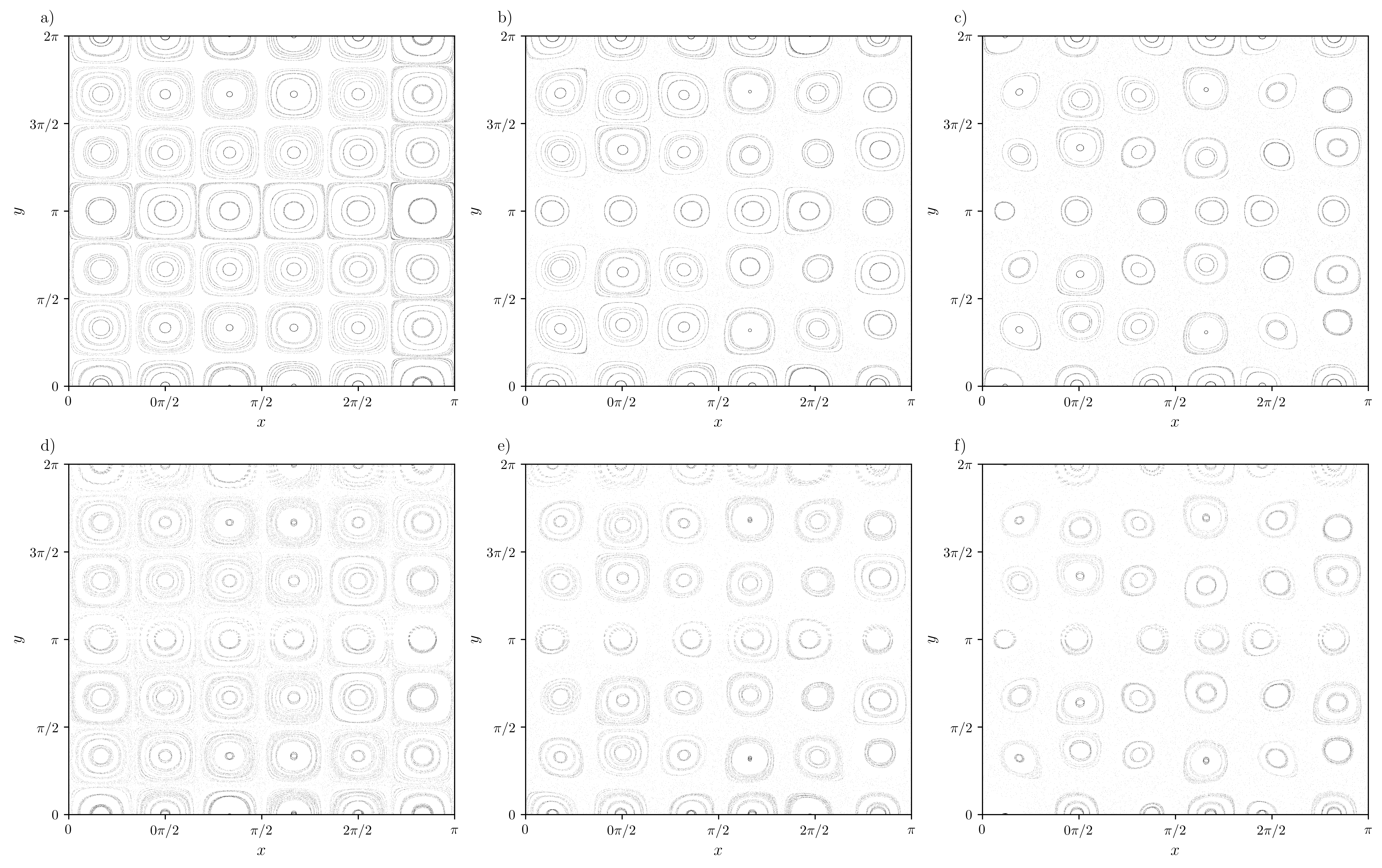}
    \caption{Phase space for two configurations: the first row corresponds to $A_3 = 0$, and the second row to $A_3 = 0.1$. Columns represent different $A_2$ values: (a,d) $0.0$, (b,e) $0.5$, and (c,f) $0.9$.}
    \label{fig:Phase_Space}
\end{figure*}

Considering two and three modes in the Hamiltonian equation \eqref{eq:H_Final}, we evolved the canonical equations \eqref{eq:vx} and \eqref{eq:vy} numerically from $t_i = 0$ to $t_f = 10^4$. A Poincaré map was constructed by sampling $(x,y)$ values at times equal to integer multiples of the period $T = 2\pi/\omega_1$. The system parameters are defined as follows:
\begin{align*}
    k_{x,1} &= 5,      & k_{y,1} &= 6,    & \omega_1 &= 2,   & A_1 &= 1.0, \\
    k_{x,2} &= -3,     & k_{y,2} &= -3,   & \omega_2 &= 2,   & A_2 &\in [0, 1.0], \\
    k_{x,3} &= -2,     & k_{y,3} &= -3,   & \omega_3 &= 2,   & A_3 &= 0.1.
\end{align*}
These values were chosen to reflect realistic plasma wave dynamics\cite{hellerCorrelationPlasmaEdge1997,batistaNonlinearThreemodeInteraction2006}, with $A_2$ (the amplitude of the second wave) serving as a tunable parameter in the range $[0.0, 1.0]$. The integrable case $A_2 = A_3 = 0$ is shown in Fig.~\ref{fig:Phase_Space}(a), where particle motion is confined to bounded cells.

By using a Poincaré map, we observe the patterns of particle motion by leveraging the periodicity of the stroboscopic mapping. This is achieved by sampling the coordinates $(x, y)$ for each particle at times $t_n = nT$, where $T = 2\pi/u_1$ is the period of the parent wave. Figure~\ref{fig:Phase_Space} illustrates the resulting maps for different values of $A_2$. Analysis reveals that increasing the control parameter $A_2$ triggers the destruction of invariant curves near displaced elliptic points, leading to a consequent enlargement of chaotic orbits. The persistence of periodic orbits at higher $A_2$ values arises from the choice of $k_x$ and $k_y$, which correspond to smaller wavelengths. These wavelengths slow the destruction of KAM tori, preserving quasi-regular motion in localized regions.

On the other hand, the addition of the third wave with $A_3 = 0.1$ (Fig.~\ref{fig:Phase_Space}(d)) breaks the integrability of the system. In this case, only the interior of the cells retains quasiperiodic motion, unlike the two-wave scenario studied earlier \cite{mathiasFractalStructuresChaotic2017}. The inclusion of the third wave accelerates the destruction of quasiperiodic islands due to the cumulative effect of multiple perturbations in the integrable system. This results in slightly more chaotic orbits in phase space and enhances particle transport, a phenomenon we will analyze in detail later.

\section{Escape Basins}
\label{sec:escape_basins}

The dynamics presented in Fig.~\ref{fig:Phase_Space} are aperiodic but not uniformly chaotic due to the intricate structure of the underlying manifolds. This non-uniformity in chaotic motion can be quantified through the geometry of escape basins. Here, "escape" is defined as particles exiting the \( L_x \times L_y \) domain, which physically corresponds to migration into distinct plasma regions: the plasma-wall boundary for \( x > \pi \) or the plasma core for \( x < 0 \). Initial conditions not residing in chaotic regions (e.g., those within periodic islands) produce non-escaping orbits and are excluded from exit basins.

\begin{figure*}[h!]
    \centering
    \includegraphics[scale=0.5]{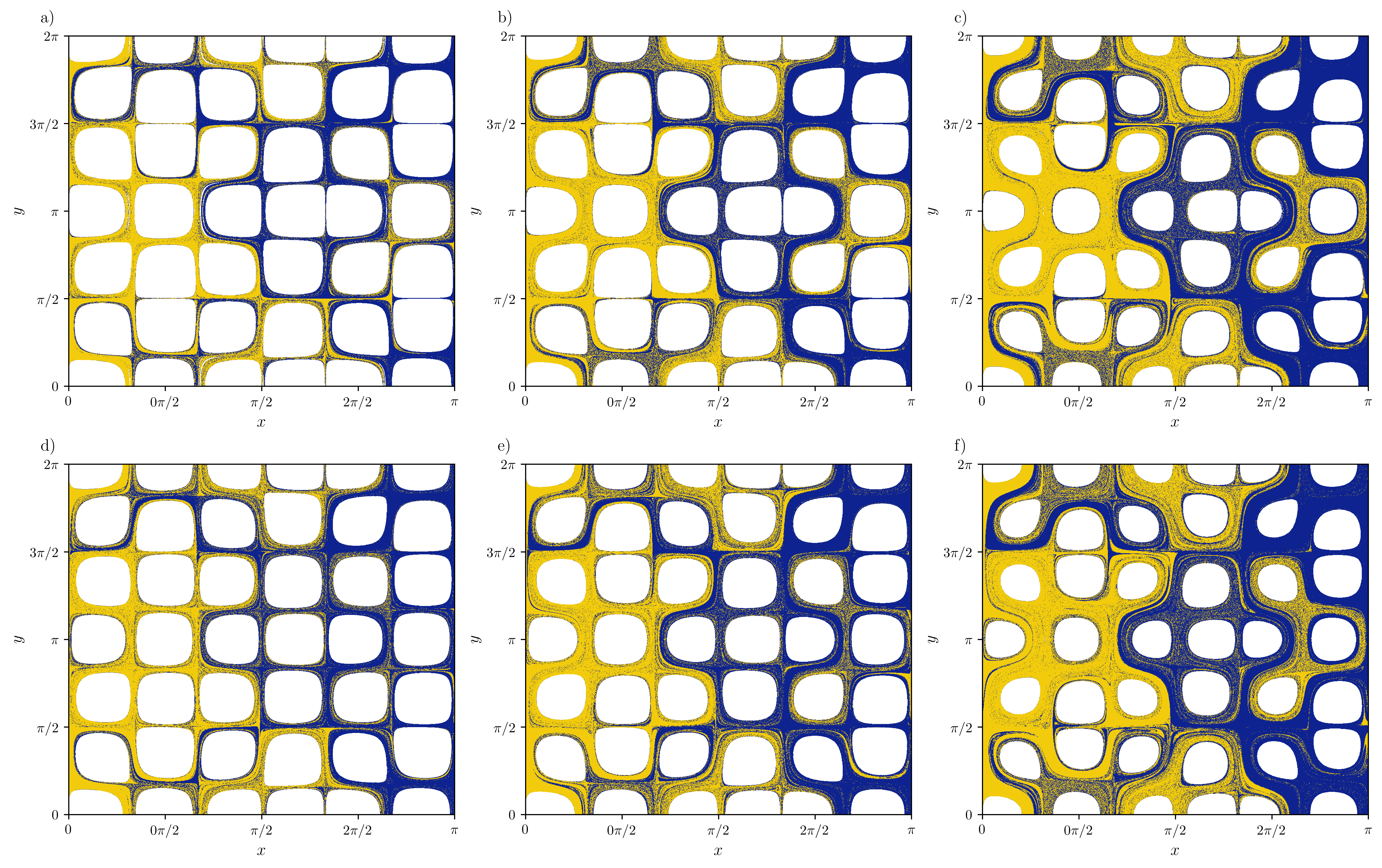}
    \caption{Escape basins for the core ($x < 0$, yellow regions) and the wall ($x > \pi$, blue regions). The first row corresponds to the system with only the second wave ($A_3 = 0$), while the second row includes the third wave ($A_3 = 0.1$). The perturbation amplitudes of the second wave are: (a,d) $A_2 = 0.3$, (b,e) $A_2 = 0.5$, and (c,f) $A_2 = 0.9$.}
    \label{fig:Escape_basins}
\end{figure*}

\begin{figure*}[h!]
    \centering
    \includegraphics[scale=0.7]{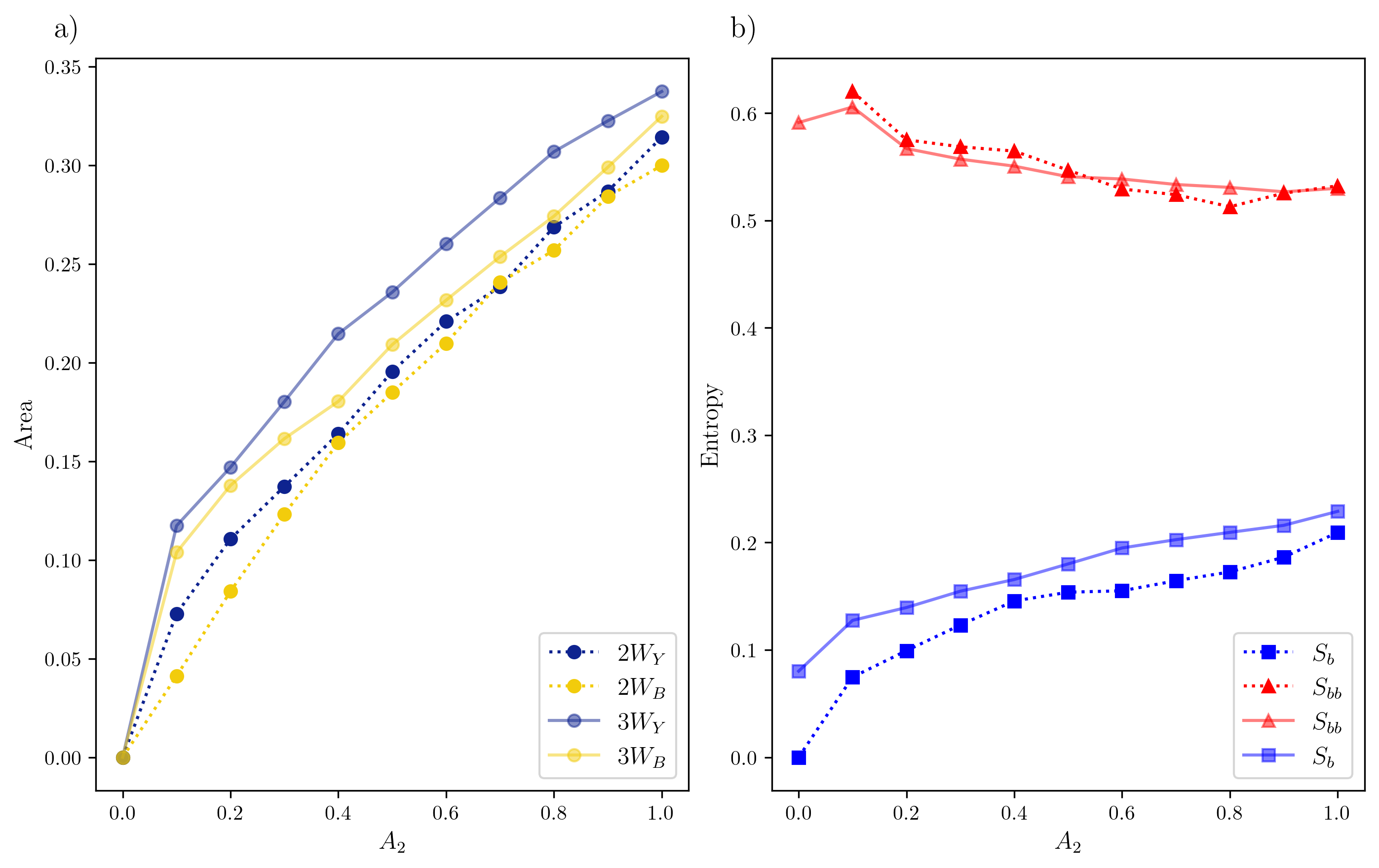}
    \caption{Analysis for two-wave ($2W$, dotted line) and three-wave ($3W$, solid line) systems across varying second-wave perturbation amplitudes $A_2$. Panel (a) shows the fraction of the occupied area, while panel (b) displays the basin entropy and basin boundary entropy for both cases.}
    \label{fig:Area_Entropy}
\end{figure*}

Fig.~\ref{fig:Escape_basins} shows the exit basins for the wall (blue-colored regions) and the core (yellow-colored regions). Initial conditions that never escape are represented as blank areas. The exit basins exhibit an intricate structure for all values of the control parameter $A_2$. The comparable size of the basins across most $A_2$ values indicates no preferential escape direction (Fig.~\ref{fig:Area_Entropy}(a)). However, this symmetry can be broken by modulating the equilibrium potential profile $\Phi_0(x)$, as demonstrated in prior work \cite{vianaFractalBoundariesChaotic2017}.  

The inclusion of the third wave with a small amplitude is linked to a physically interesting phenomenon in plasma physics \cite{batistaNonlinearThreemodeInteraction2006,hasegawaPseudothreedimensionalTurbulenceMagnetized1978}. While its effects are hardly perceived by direct comparison of exit basins in Fig.~\ref{fig:Escape_basins}, we quantify the basin areas by calculating the fraction of basin points relative to the total number of grid points dividing the phase portrait. By comparing the sizes of the two basins for each case (Fig.~\ref{fig:Area_Entropy}a), it is evident that the third wave causes more particles to escape, with an increase of approximately $3\%$. The positive trend in all escape basins with increasing $A_2$ can be attributed to the enlargement of chaotic orbits. This expansion increases the number of initial conditions available for trajectories that eventually escape to either boundary.

The escape basins in our system are so densely intertwined that the fractal dimension of their boundary approximates that of the entire phase plane. Consequently, the basin boundary exhibits a structure analogous to space-filling curves, such as the Peano or Hilbert curves. Notably, this fractal dimension remains nearly invariant under variations of the control parameter $A_2$. As a result, analyzing the basin boundary dimension does not provide meaningful insight into the fractality of the escape basins or their degree of final-state sensitivity across different $A_2$ values.  

To quantify the effects of different exit configurations, we employ basin entropy analysis \cite{dazaBasinEntropyNew2016,haerterBasinEntropyWada2023,dazaClassifyingBasinsAttraction2022}. This method partitions the escape region into a finite grid of $M$ boxes, each of size $\varepsilon$, and calculates the escape probabilities under the assumption of equiprobability among all boxes:
\begin{equation}
  p_e = \frac{n_e}{\varepsilon^2},
  \label{eq:pe}
\end{equation}
where $e \in [1, N_A]$ denotes the exit index, and $N_A$ is the total number of exits. The Shannon entropy for the $i$-th cell is defined as:
\begin{equation}
    \label{eq:entro}
    S_i = -\sum_{e=1}^{N_A} p_e \log p_e.
\end{equation}

Since this entropy is extensive, the total basin entropy of the system is obtained by averaging over all $M$ boxes:
\begin{equation}
  S_b = \frac{1}{M} \sum_{i=1}^{M} S_i,
  \label{eq:Sb}
\end{equation}
referred to as the basin entropy.

For boxes containing only one attractor, the Shannon entropy $S_i$ vanishes. By excluding these boxes, we define the basin boundary entropy:

\begin{equation}
  S_{bb} = \frac{1}{N_b} \sum_{i=1}^{M} S_i,
  \label{eq:Sbb}
\end{equation}

where $N_b$ is the number of boxes with mixed attractors. This measure is more sensitive to phase-space topology and the proximity of multiple attractors.

These two metrics can analyze the complexity, mixing, and uncertainty of initial conditions within escape basins or basins of attraction. They are closely linked to established metrics like the fractal dimension of attractors and the uncertainty exponent \cite{dazaClassifyingBasinsAttraction2022}. When $S_b \rightarrow 0$, nearly all initial conditions within a small box of size $\varepsilon^2$ converge to the same exit, reflecting minimal uncertainty and higher trajectory predictability. Conversely, when $S_b \rightarrow \ln N_A$, initial conditions within the box are distributed equally among all exits, indicating maximal uncertainty and unpredictability.

Fig.~\ref{fig:Area_Entropy}(b) displays the basin entropy and basin boundary entropy for both analyzed cases. As expected, the results show that the basin entropy increases with larger values of the second wave’s amplitude ($A_2$). This increase arises from enhanced chaotic behavior in the system, which amplifies the fractality and structural complexity of the basins.

When comparing the basin entropy $S_b$, the three-wave case exhibits higher values for all $A_2$, including a nonzero $S_b$ at $A_2 = 0.0$. This nonzero entropy arises from the breaking of invariant curves due to the third wave, which ensures the system always exhibits escape basins. This is further corroborated by the presence of basin boundary entropy ($S_{bb}$) even at $A_2 = 0.0$. The basin entropy values are consistently higher in the presence of the third wave, demonstrating that even a small additional perturbation disrupts basin mixing and amplifies uncertainty.

The $S_{bb}$, on the other hand, presents a lower value when the third wave is added. This is counterintuitive but expected: while the third wave increases mixing in the system (producing more boundary points), the basin boundary entropy $S_{bb}$ decreases because the averaging denominator $N_b$ (number of mixed-attractor boxes) grows faster than the sum of entropies. The negative trend in $S_{bb}$ arises as the escape regions expand. With both core and wall basins growing proportionally, the occupied area increases, but the uncertainty grows only marginally. Meanwhile, the rise in $N_b$ reflects a larger number of mixed-attractor regions, diluting $S_{bb}$ despite the heightened structural complexity.

\begin{figure*}[h!]
    \centering
    \includegraphics[scale=0.5]{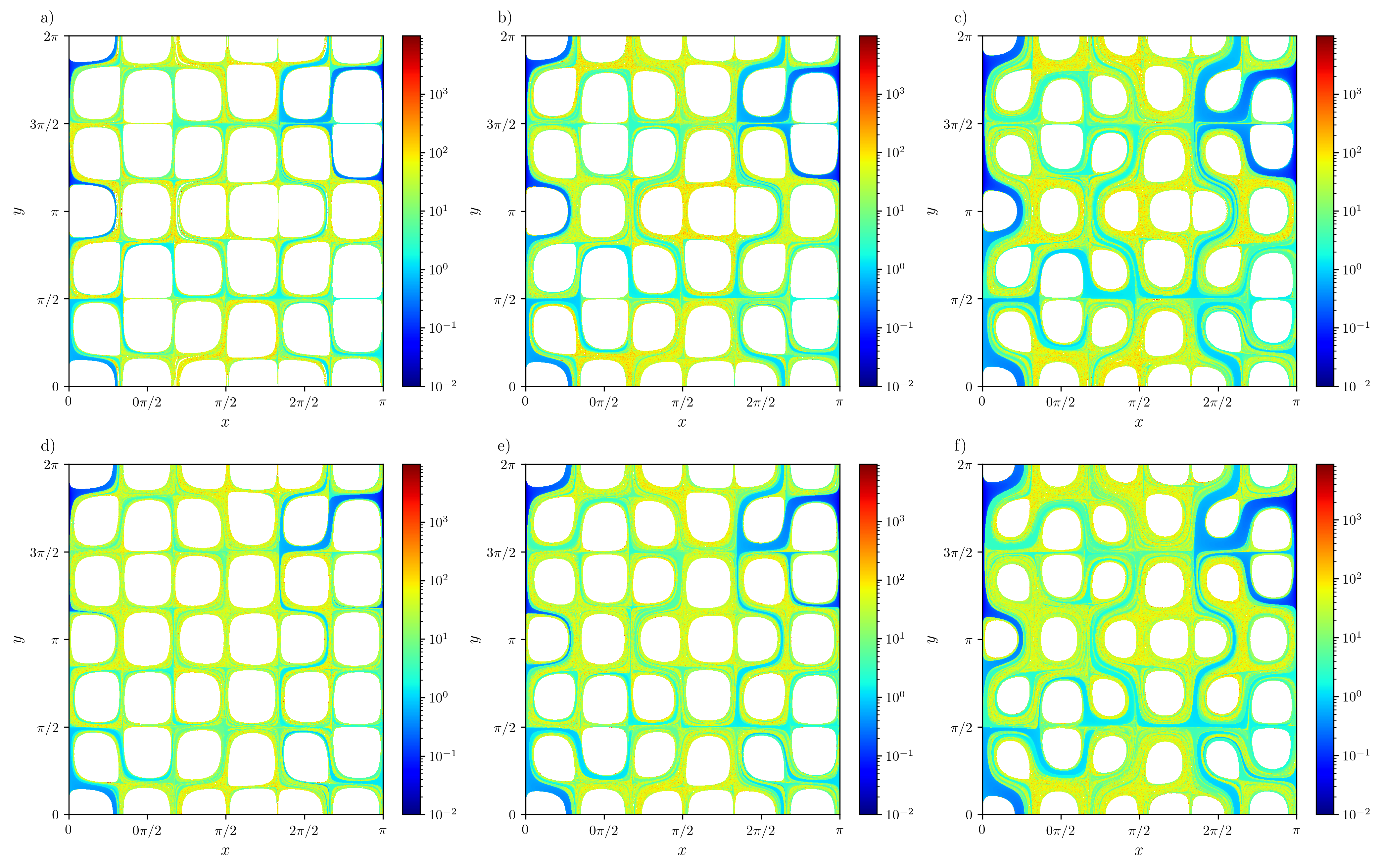}
    \caption{Escape time for particles in two configurations: the first row corresponds to the system with only the second wave ($A_3 = 0$), while the second row includes the third wave ($A_3 = 0.1$). Columns represent different second-wave perturbation amplitudes: (a,d) $A_2 = 0.3$, (b,e) $A_2 = 0.5$, and (c,f) $A_2 = 0.9$. Warmer colors indicate longer escape times, while cooler colors denote faster escapes.}
    \label{fig:Escape_time}
\end{figure*}

Furthermore, it is also possible to analyze the escape time of particles for different second-wave amplitudes. Fig.~\ref{fig:Escape_time} illustrates this relationship, where warmer colors correspond to longer escape times and cooler colors indicate faster-escaping particles. The fast-escape regions correlate with smooth basin boundaries, a consequence of unstable manifolds in the system and reduced divergence among initial conditions. Conversely, regions with prolonged escape times are associated with sticky orbits—trajectories that remain transiently trapped near periodic orbits before escaping. These orbits also contribute to the uncertainty regions observed in escape basins.

\section{Particle Transport}
\label{sec:particle_transport}

To analyze the transport dynamics induced by the presence of two and three waves, we simulated an ensemble of passive particles initially distributed within the same region studied for escape basin characterization.  

We employed diagnostic measures to quantify how distinct wave configurations influence particle transport in tokamaks. Specifically, we calculated two metrics: the mean square displacement (MSD), which provides a global characterization of chaotic transport by quantifying the spatial spread of particles over time, and the total displacement, which captures the inhomogeneous nature of transport due to spatially varying flux intensities across plasma regions.  
 
\subsection{Mean Square Displacement}

For a statistical characterization of transport, we computed the mean square displacement (MSD):
\begin{equation}
    \langle\sigma(t)^2\rangle = \frac{1}{N}\sum_{i=1}^N \qty|\vb{x}_i(t) - \vb{x}_i(0)|^2,
    \label{eq:MSD}
\end{equation}
where $N$ denotes the total number of particles. The temporal scaling exponent $\langle\sigma(t)^2\rangle \propto t^\alpha$ classifies the transport regime as subdiffusive ($\alpha < 1$), normal diffusion ($\alpha = 1$), superdiffusive ($1 < \alpha < 2$), or ballistic ($\alpha = 2$).

\begin{figure}[h!]
    \centering
    \includegraphics[width=0.5\textwidth]{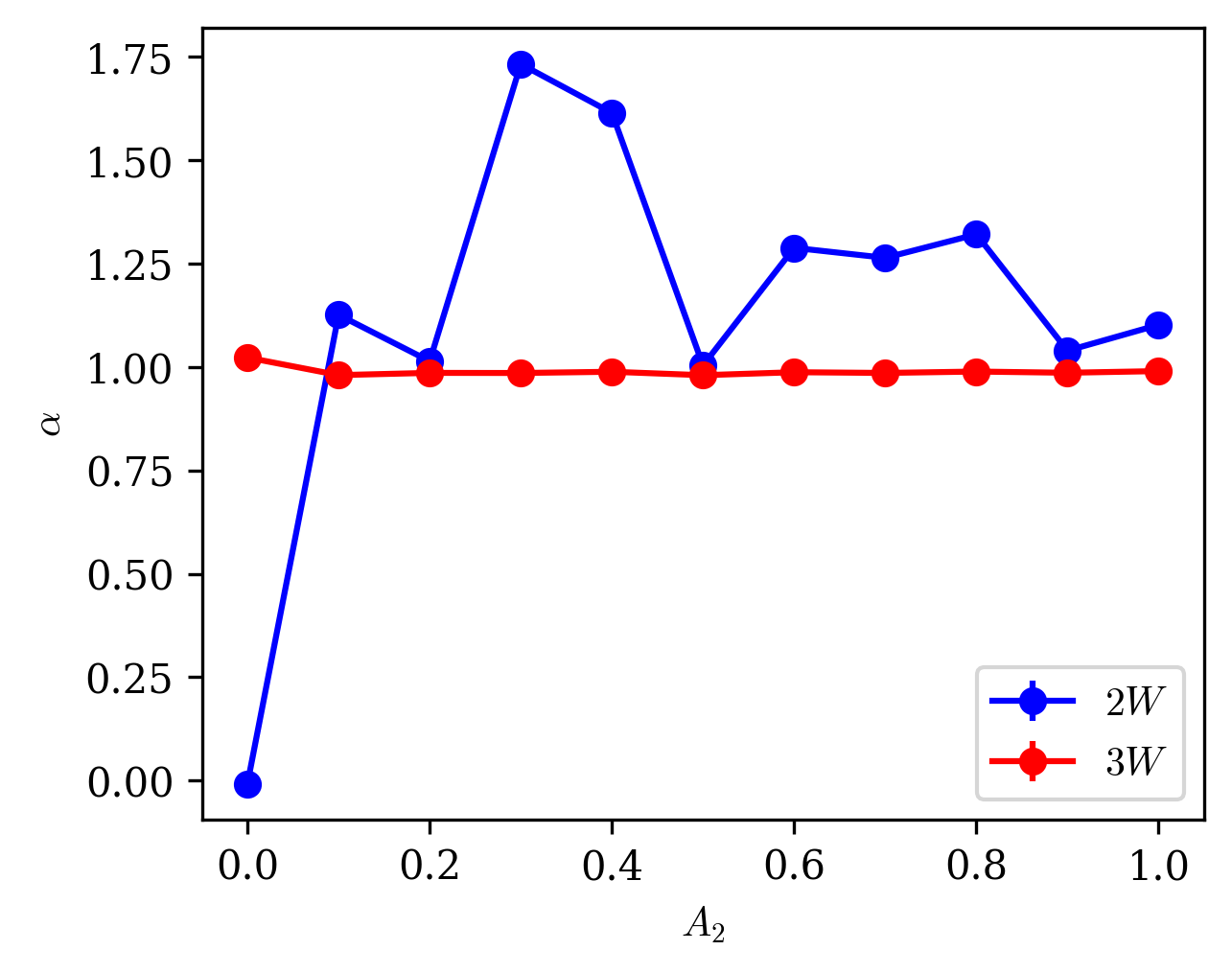}
    \caption{Mean square displacement (MSD) of particles for two- and three-wave systems at multiple values of the second wave perturbation amplitude ($A_2 \in [0.0, 1.0]$). The results highlight the confinement effect induced by the third wave and the nonlinear dependence of displacement on $A_2$ in the two-wave case.}
    \label{fig:MSD}
\end{figure}

Our analysis reveals two distinct regimes of anomalous transport, characterized by different scaling exponents $(\alpha)$, which emerge as a direct consequence of the waves' presence. Despite similar escape characteristics, two- and three-wave systems induce markedly different transport dynamics. In the two-wave case, particles exhibit enhanced poloidal transport, leading to spatially heterogeneous high-flux regions. Conversely, the addition of a third wave suppresses these localized transport channels due to resonant interactions, resulting in nearly constant diffusion across parameter space and effective poloidal confinement. This resonance mechanism homogenizes transport fluxes, reducing overall system transport compared to the two-wave scenario.
The Figure \ref{fig:MSD} summarizes the MSD scaling across the parameter space for the two and three wave, revealing a clear distinction between two groups: one corresponding to two waves system (Blue curve) and the three waves (Red curve).

The observed diffusion exponent provides a quantitative measure of the transport regimes and their underlying mechanisms. The anomalous diffusion regime observed for the two-wave system suggests the formation of highly correlated regions where particles exhibit persistent directional coherence. This behavior mirrors coherent structures in plasma turbulence, such as zonal flows or streamers, which efficiently channel particles along preferential pathways. This phenomenon stems from the two waves generating localized high-transport channels within the chaotic sea. However, as the second wave amplitude $(A_2)$ increases, the amplified chaotic dynamics disrupt these channels—reducing periodic orbits and suppressing transport efficiency.

In contrast, the normal diffusion regime under three waves indicates a structured phase space without the formation of high-transport regions. Here, directional coherence—which drives localized transport—is suppressed by the third wave. This suppression arises because the third wave introduces resonant interactions that amplify background stochasticity, trapping particles in transient coherent structures or inducing frequent scattering. These mechanisms disrupt anomalous-diffusion pathways, homogenizing transport and preventing large-scale particle flux.

\subsection{Displacement}

The total displacement of particles ($\Delta$) is defined as the Euclidean distance between their initial and final positions:
\begin{equation}
    \Delta = \sqrt{(\Delta x)^2 + (\Delta y)^2},
\end{equation}
providing critical insights into both the magnitude and directionality of transport. Unlike the mean square displacement (MSD), which quantifies particle dispersion, the total displacement highlights net migration of particles. By analyzing $\Delta$ alongside directional displacements ($\Delta x$, $\Delta y$), we observe two distinct regimes identified in the MSD analysis: localized high-transport regions in the two-wave system, where particles exhibit coherent motion, and diffuse transport in the three-wave system, where the third wave disrupts directional coherence.

The total displacement $\Delta$ is consistent with the scaling exponent $\alpha$ in the MSD analysis. By comparing displacements across spatial regions, we identified areas of enhanced transport. These regions correlate with chaotic dynamics in phase space, while suppressed transport aligns with quasi-regular orbits trapped near invariant structures.
\begin{figure*}[h!]
    \centering
    \includegraphics[scale=0.5]{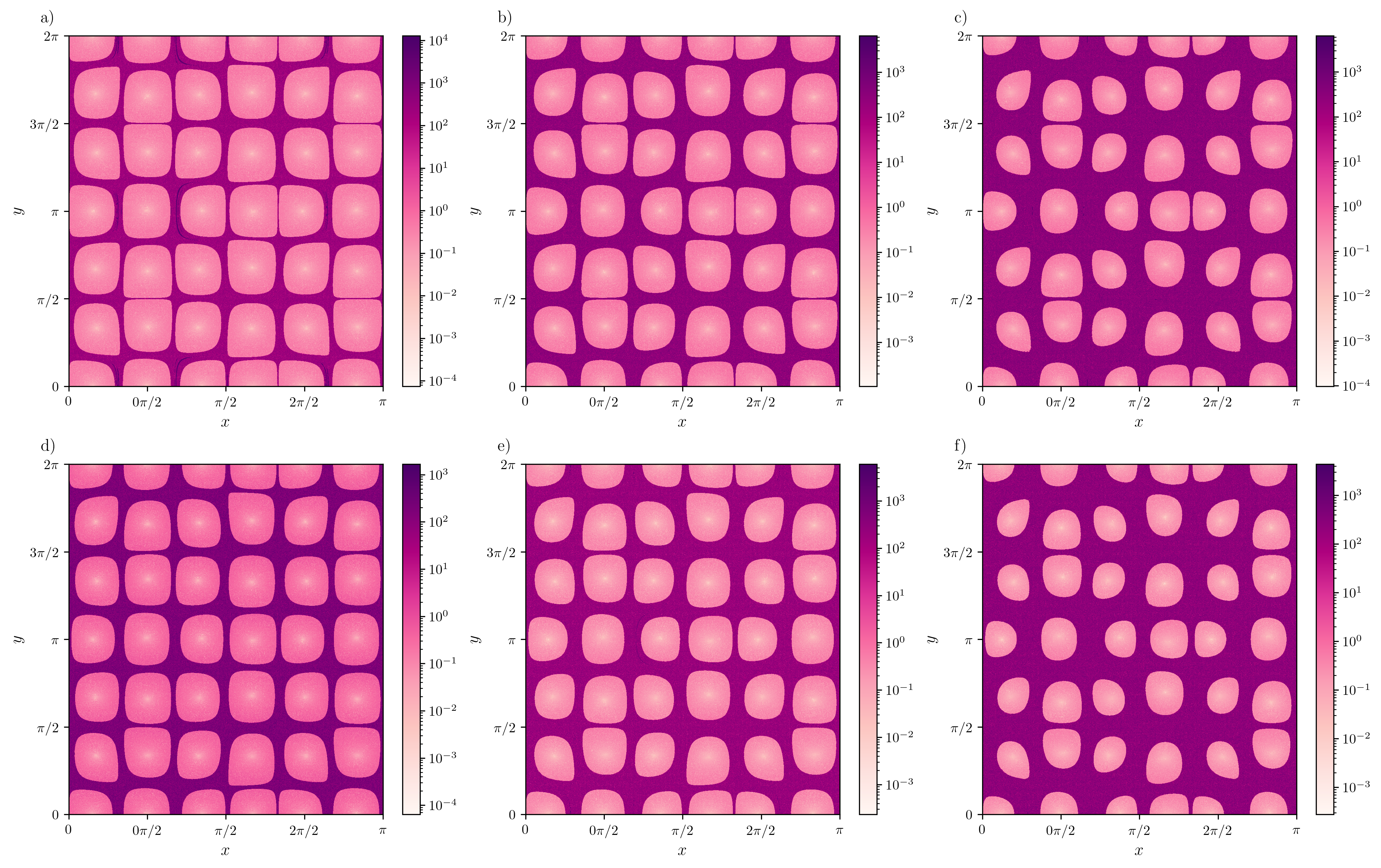}
    \caption{Total displacement ($\Delta$) of advected particles. The first row corresponds to the two-wave system ($A_3 = 0$), and the second row includes the third wave ($A_3 = 0.1$). Columns represent different second-wave perturbation amplitudes: (a,d) $A_2 = 0.3$, (b,e) $A_2 = 0.5$, and (c,f) $A_2 = 0.9$.}
    \label{fig:Tot_Displ}
\end{figure*}

Figure~\ref{fig:Tot_Displ} illustrates the displacement patterns observed for different wave configurations. The first row corresponds to the two-wave system ($A_3 = 0$), while the second row shows the three-wave system ($A_3 = 0.1$). Columns are ordered by increasing second-wave perturbation amplitude: $A_2 = 0.3$, $A_2 = 0.5$, and $A_2 = 0.9$ (left to right).

The displacement metric reinforces the separation of transport regimes observed in Fig.~\ref{fig:MSD}. For the two-wave system (first row of Fig.~\ref{fig:Tot_Displ}), the total displacement reveals distinct dynamics: particles travel farther when the scaling exponent $\alpha$ is higher (e.g., $A_2 = 0.3$), with localized high-transport regions near $(1.0, \pi)$. However, these regions are suppressed as the wave amplitude ($A_2$) increases. Conversely, in cases where $\alpha \approx 1$, the displacement exhibits similar behavior across parameters, with no high-transport structures forming. Chaotic regions instead display isotropic displacement due to frequent scattering, which suppresses net migration by trapping particles in transient structures. This scattering arises from both the increased second-wave amplitude and the presence of the third wave, hindering the formation of structured transport pathways.

\section{Conclusions}
\label{sec:conclusions}

This work investigates the role of drift-wave modes in governing particle escape and transport dynamics in tokamak edge plasmas. By comparing systems with two and three waves, we demonstrate how the introduction of a third wave modifies uncertainty and transport regimes. 

The addition of a third wave increases the overall particle escape by approximately $3\%$. The underlying dynamical changes also lead to an increase in the basin entropy $S_b$, reflecting greater unpredictability in particle destinations. Concurrently, the basin boundary entropy $S_{bb}$ is generally reduced, which can be attributed to how this metric averages uncertainty over an increasing number of mixed boundary boxes. This suggests that while overall escape is enhanced and general phase-space unpredictability $(S_b)$ rises, the complexity specifically at the basin boundaries (as measured by $S_{bb}  $) shows a constrained behavior. Auxiliary diagnostics, including escape time and phase-space trajectory analyses, corroborate these results, revealing suppressed coherent transport pathways under three-wave configurations. The interplay between enhanced overall escape, increased general unpredictability, and this constrained nature of boundary complexity underscores the potential of multi-wave turbulence control for optimizing plasma confinement strategies.

The suppression effect of the third wave extends to transport dynamics. Systems with two drift waves exhibit anomalous diffusion $(\alpha > 1)$, driven by high transport regions that channel particles along preferential pathways. In contrast, the third wave disrupts these pathways through resonant scattering, restoring normal diffusion $(\alpha \approx 1)$ and homogenizing transport fluxes. This duality highlights the delicate balance between order and chaos in multi-wave plasmas, with profound implications for confinement optimization and divertor design. By suppressing anomalous transport while maintaining confinement efficiency, controlled wave interactions offer a pathway to stabilize particle motion and reduce heat loads on reactor walls.

These results bridge fundamental plasma turbulence studies to practical confinement challenges. The observed fractal structures in escape basins, coupled with the tunable transition between anomalous and normal diffusion, suggest that multi-wave configurations could be engineered to tailor transport properties. Future work should explore scaling these findings to realistic reactor geometries and experimental validation of entropy-based control strategies.

\section*{Acknowledgments}
This work has been supported by grants from the Brazilian Government Agencies CNPq and CAPES. P. Haerter received partial financial support from the following Brazilian government agencies: CNPq (140920/2022-6), CAPES (88887.898818/2023-00). E.D.Leonel acknowledges support from Brazilian agencies CNPq (No. 301318/2019-0, 304398/2023-3) and FAPESP (No. 2019/14038-6 and No. 2021/09519-5). R. L. Viana received partial financial support from the following Brazilian government agencies: CNPq (403120/2021-7, 301019/2019-3), CAPES (88881.143103/2017-01).

\bibliographystyle{elsarticle-num}
\bibliography{mybibfile}
\end{document}